# Exchange-Coupled Spins for Robust High-Temperature Qubits

Aniruddha Chakraborty[1], Md. Fahim F. Chowdhury[1], Mohamad Niknam[2,3], Louis S. Bouchard[2,3], Jayasimha Atulasimha[1, 4, 5]✉

[1]Department of Mechanical and Nuclear Engineering, College of Engineering, Virginia Commonwealth University, Richmond, VA 23284-3068, USA.

[2]Department of Chemistry and Biochemistry, University of California Los Angeles, 607 Charles E. Young Drive East, Los Angeles, CA 90095-1059, USA.

[3]Center for Quantum Science and Engineering, UCLA, Los Angeles, CA 90095-1059, USA.

[4]Department of Electrical and Computer Engineering, College of Engineering, Virginia Commonwealth University, Richmond, VA 23284-3068, USA.

[5]Department of Physics, College of Humanities and Sciences, Virginia Commonwealth University, Richmond, VA 23284-3068, USA.

✉email: jatulasimha@vcu.edu.

## Abstract
We show that Heisenberg exchange interactions between the neighboring spins comprising an ensemble spin qubit (E-qubit) can act as an intrinsic error mitigator, increasing gate fidelity even at high temperatures. As an example, the fidelity of a $\pi_x$ gate applied to E-qubits above $1\,K$ was studied by tuning the ferromagnetic exchange strength to show an exchange coupled E-qubit exhibits higher fidelity than a single-spin based qubit. We also investigate the coherence properties of E-qubits and find that the coherence time of an E-qubit extends linearly with the number of spins in the ensemble. This suggests that exchange interactions effectively suppress decoherence induced by thermal noise, achieving a coherence time greater than $1\,ms$ at 1 K with an ensemble of only seven spins. Additionally, the ferromagnetic isotropic exchange prevents fidelity loss induced by spatial field gradients/inhomogeneity in Zeeman and/or control fields. Therefore, exchange-coupled spin qubits could enable fault-tolerant quantum operations and long-coherence times at elevated temperatures (>1 $K$).

## Introduction

In current quantum technology, there are many proposals for physical realizations of qubits, including trapped ions[1], nuclear spins[2–4], superconducting circuits[5], quantum dots[6,7], semiconductor spins qubits[8], NV center in diamonds[9,10]. Among these technologies, spin qubits were first experimentally realized and are still regarded as a benchmark system for exploring quantum control and quantum information. The main challenge in building quantum computers using spin qubits is controlling and detecting single spins, because the magnetization of an individual spin is extremely small. Ensemble qubits (E-qubits) consist of many particles/spins where the average evolution of the ensemble provides a more robust platform for performing quantum operations. Those use a novel way for averaging the spin dynamics and are promising candidates for large-scale quantum devices[11]. As electric field noise has a negligible effect on the spin degree of freedom, spin qubit has the advantage of having a long coherence time[12]. Moreover, qubit fabrication with modern lithography techniques and other semiconductor manufacturing processes paves the way for the development of spin qubits as a scalable technology.

Spin qubits in quantum dots face several challenges, including decoherence induced by hyperfine interaction, charge noise, and thermal noise. These noise sources restrict the qubit operation to the milli Kelvin temperature range[13]. The accumulated error still needs to be corrected by quantum error circuitry or protocol to prevent this exponential error accumulation, increasing the quantum operation's complexity and affecting scalability[14].

The E-qubit offers the presence of multiple copies of spin states, which would allow for implementation of local probes for quantum control and detection of single E-qubits[11]. Moreover, an ensemble of spins offers other favorable features, such as a simplified fabrication process and enhanced quantum gate operations[15,16]. Large physical volumes for qubits as an ensemble of spins can make the fabrication process less demanding than single-spin / atom qubits, where accurate implantation of defects is a technological limitation. In recent studies, it has been proposed that, with proximally controlling uncoupled spin ensemble with nanomagnets[17] or with magnetic Skyrmions[18], it is possible to achieve a high fidelity $\pi_x$ gate.

Here, we demonstrate that an E-qubit (for example, ensemble of TM or RE ions), acting as a single qubit entity due to ferromagnetic exchange coupling, are robust to the spin decoherence due to thermal noise. In this setup, the $\pi_x$ gate can be implemented with fidelity of more than $99\,\%$, at $6\,K$, while at $1\,K$ a fidelity of $\sim99.9\,\%$ can be achieved. Moreover, coherence time can be extended to $1.1\,ms$ at $2\,K$ in an ensemble qubit consisting of just seven spins. Finding a "hot qubit", which can be operational at high temperature is a challenging problem which can have a significant impact on the field of quantum information[19].

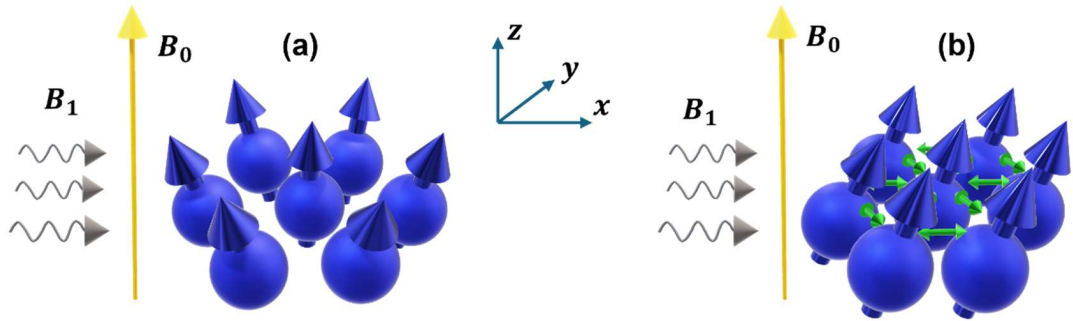

**Figure 1: Comparison schematics of spins orientation in ensemble under magnetic resonance.** **(a).** Uncoupled spins in the ensemble de-phase easily. **(b)** Exchange coupled spins in the ensemble have minimum dephasing. Exchange couplings between neighboring spins are indicated by double headed green arrows. Here, yellow arrow indicates Zeeman field ($B_0$), and wavy grey arrow indicates microwave applied as a control field ($B_1$).

## Results and Discussion

An E-qubit can be conceptualized as a few exchanged coupled spins embedded in a highly localized region of a substrate. Typically, a combination of static and locally applied time-dependent magnetic fields is used to control the qubits at resonant condition[20]. At low temperatures, this system can be initialized in a fully polarized state either spin-up or spin-down, by applying Zeeman field perpendicular to the plane of the $2D$ ensemble lattice. These spins can be realized through the deposition of $3d$ or $4f$ magnetic atoms on an oxide interface leveraging the advantage of having perpendicular magnetic anisotropy (PMA) induced by spin orbit coupling as observed in case of a single Co spin on MgO[21]. The implementation of a $\pi_x$ gate is achieved by applying a uniform microwave field on the ensemble in the transverse direction-perpendicular to the static magnetic field axis, thus inducing coherent rotation of spins and enabling ensemble qubit manipulation.

For this simulation study, the parameters used are listed in **Table 1**.

**Table 1: List of parameters used in the simulation.**

| Parameters | Value |
|---|---|
| Global bias magnetic Field, $B_0$ | $71.36\ mT$ |
| Microwave field amplitude, $B_1$ | $1.78\ mT$ |
| Frequency of microwave field, $f$ | $2\ GHz$ |
| Spins gyromagnetic ratio, $\gamma_e$ | $1.7609 \times 10^{11}\ rad.s^{-1}.T^{-1}$ |
| Magnetic permeability, $\mu_0$ | $4\pi \times 10^{-7}\ N.A^{-2}$ |
| Magnetic moment of electron, $\mu_e$ | $9.274 \times 10^{-24}\ J.T^{-1}$ |

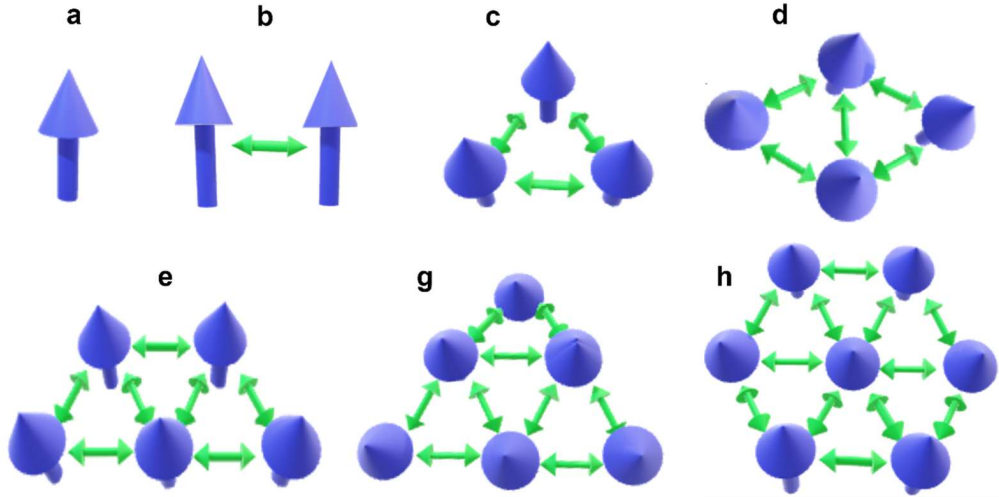

**Figure 2. Ensemble arrangements with different number of spins** Spins are represented by blue arrows in a $2D$ triangular arrangement. The ferromagnetic exchange interaction is indicated by the green double-headed arrow acting between neighboring spins.

**In** this study, thermal noise is modeled as stochastic random valued fluctuations, as described in equation (6). We investigate the dynamics of different electron spin ensembles by applying a $\pi_x$ gate to the E-qubit. By describing the evolution of spins in the lab frame, we observe that, despite the presence of thermal noise and the spatial field gradient effects, a robust high-fidelity quantum gate realization is possible. When the spins are positioned in proximity, their wavefunctions overlap, inducing an isotropic ferromagnetic exchange interaction. Uncoupled spins experience dephasing, whereas the exchange-coupled spins maintain high coherence during Rabi rotation upon applying a $\pi_x$ gate, as shown schematically in Figure 1.

For the simulation study, we consider up to seven exchange-coupled spins arrangements. The spins are assumed to be closely packed on a 2D lattice plane, forming triangular arrangement as shown in Figure 2.

Under ideal conditions, with no thermal noise, a $\pi_x$ gate can be applied by using microwave field with an amplitude $B_1 = 1.78\ mT$ and a modulation frequency of $\frac{(\gamma_e B_1)}{2\pi} = 50\ MHz$ with pulse duration $20\ ns$. Consider an exchange-coupled spin ensemble consisting of six spins initialized in the ground state $|0\rangle$ with a static magnetic field $B_0$ along $z$-direction. In this case, a resonant microwave field with amplitude $B_1$ is applied along $x$ direction for the $\pi_x$ gate operation. All spins undergo a collective coherent rotation in the presence of thermal noise. The result of $\pi_x$ gate operation on a six-spin ensemble is compared with the $\pi_x$ gate operation on a single spin in Figure 3. Since the ensemble provides statistical information about overall qubit states, it requires $N$-fold fewer quantum operations (where $N$ is the number of spins in ensemble) compared to single-spin qubit. For instance, the average spin dynamics of a single spin qubit subjected to a $\pi_x$ gate were simulated 180 times at $6\,K$. In contrast, the average spin dynamics of an E-qubit, consisting of six spins subjected to a $\pi_x$ gate, were simulated 30 times at the same temperature. These results are shown in Figure 3(a). The dynamics curves in Figure 3(a) appear smooth as the fluctuating terms are mostly averaged out, leaving the expected values across all iterative operations. Because the exchange coupling among neighboring spins reduces decoherence, the final state of the six-spin ensemble qubit is closer to the south pole of the Bloch sphere than that of a single-spin qubit, as shown in Figure 3(b).

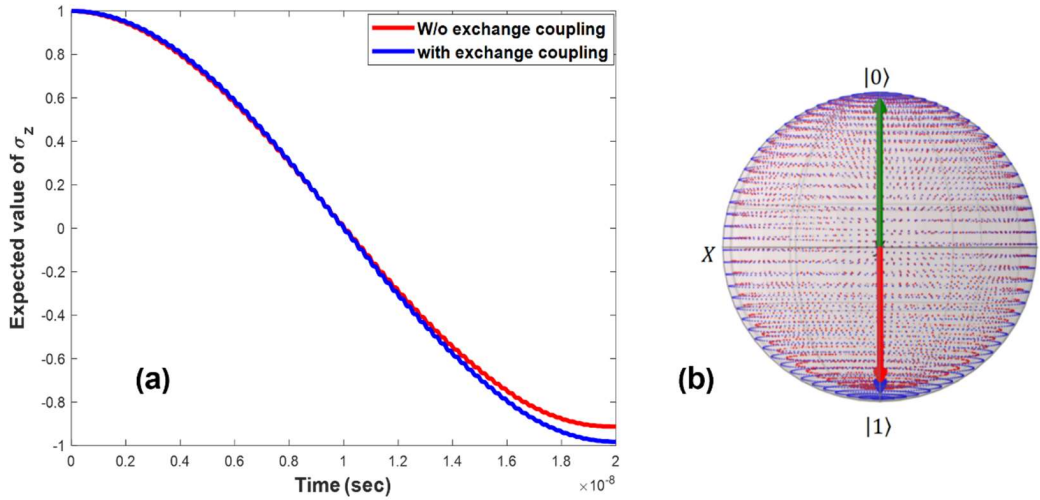

**Figure 3. Comparison of spin dynamics for a $\pi_x$ gate operation.** **(a)** The plot shows a comparison of the lab-frame spin dynamics of a single-spin qubit at $6\,K$ temperature (red curve) and the spin dynamics of a ferromagnetically coupled six-spin ensemble at the same temperature (blue curve). The projection of the expected value along the Zeeman field direction, $\rho_z$ is more negative for the six-spin ensemble compared to the single-spin dynamics, indicating the enhanced robustness of the spin ensemble. **(b)** The Bloch sphere depicts the overall trajectory of spin during the $\pi_x$ gate operation. All spinors are initialized in the ground state $|0\rangle$ (green arrow). After applying the $\pi_x$ gate, the final spinor vector of single spin qubit (red arrow) shows higher deviation from the south pole of the Bloch sphere, while the spinor of six-spins E-qubit (blue arrow) is closer to the south pole.

**Comparative analysis of $\pi_x$ gate fidelity on E-qubit.** To reduce the burden of error correction circuits in quantum information processing systems, high-fidelity gate operations are essential. The fidelity of a quantum gate is calculated by comparing the evolution of qubits under an ideal operation, with the state resulting from unitary propagator[22,23] that includes the effect of noise. First, the reduced density matrix of each individual spin in the ensemble is calculated by taking the partial trace of density matrix of E-qubit after the $\pi_x$ gate. The resulting state of each individual spin is then compared with the ideal state to determine the fidelity of each spin in the ensemble. By averaging the fidelities of all individual spins and averaging fidelities from all iterative operations, the estimated fidelity of the noisy gate for the E-qubit at a given temperature can be obtained.

The exchange coupling between the neighboring spins tends to align the spins in parallel, assisting in maintaining collective coherent rotation. To represent thermal noise as white noise, we use the Langevin term consisting of a Gaussian random field distribution with mean 0, and standard deviation of $\sqrt{\frac{2\gamma_e \mu_0 k_B T \alpha}{\mu_e \Delta t}}$, where $T$ is the temperature, $\mu_e$ is the magnetic moment of the spin, $\Delta t$ is the time step, $\alpha = 0.5$ is the damping factor[24]. For a six-spin E-qubit, the average quantum gate fidelity improves with stronger exchange coupling, as shown in Figure 4. Notably, even at $6\ K$, the $99\ \%$ fidelity threshold can be achievable for exchange coupling strength larger than $6.5 \times 10^{-2}\ J$ per site. The exchange coupling strength required to achieve $0.99\ \%$ gate fidelity for a given spin ensemble qubit increases with operating temperature as more coupling is needed to suppress the decoherence effect caused by the increased thermal noise.

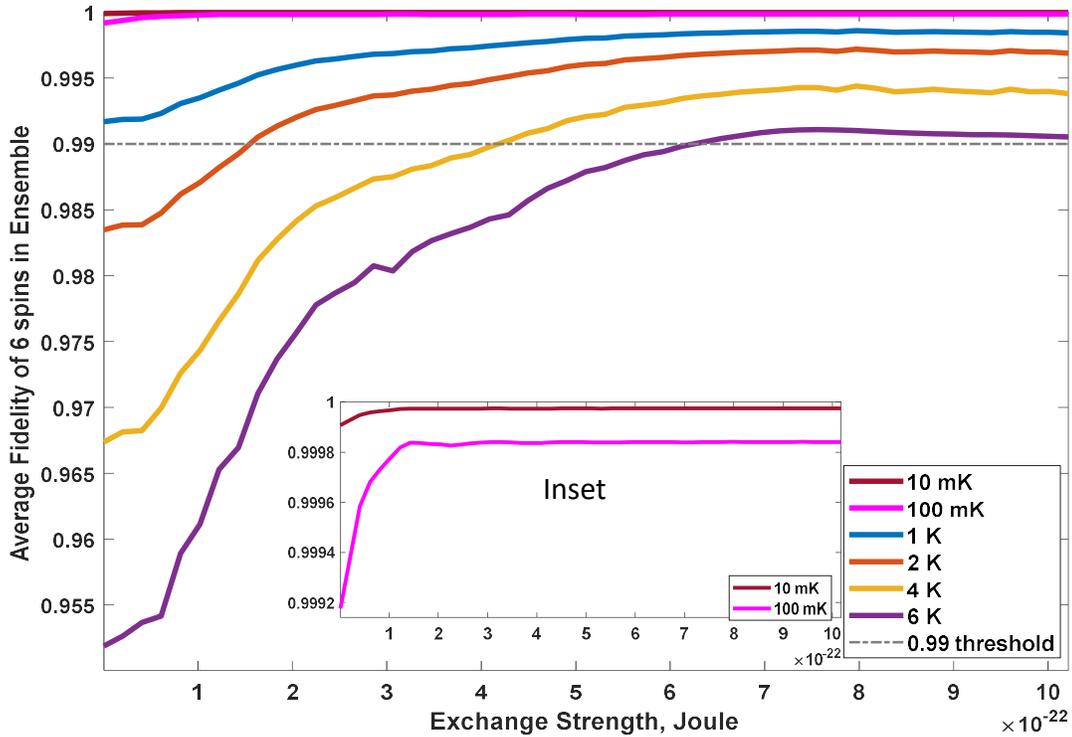

**Figure 4. Dependence of average fidelity of six-spin ensemble on exchange coupling strength.** The average fidelity of six-spin ensemble is calculated by performing $420$ iterative $\pi_x$ gate operations at different temperatures. This figure underlines the effectiveness of exchange coupling in maintaining High-fidelity during high temperature ($6\ K$) operation.

For a given exchange strength, we observe that the gate fidelity of an E-qubit depends on the number of spins in the ensemble. A larger number of spins offers the advantage of maintaining higher coherence time and fidelity, as more exchange coupling enhances the system's robustness. Moreover, in the system with a larger number of spins, the noise effects tend to average out. Figure 5 shows the $\pi_x$ gate fidelity for E-qubits of different sizes at varying temperatures, with an exchange strength of $7.15 \times 10^{-22} J$ per site between neighboring spins. As expected, for each qubit size, higher temperature results in noisier gate operation. However, we observe that as the number of spins increases, fidelity consistently increases. The fidelity plot in Figure 5 is drawn by averaging the $\pi_x$ gate fidelity for single-spin qubit over 1680 times operations while averaging the fidelity of the $\pi_x$ gate on $N$-spin E-qubit over $1680/N$ operations. At $1\ K$, the $\pi_x$ gate fidelity for a single spin qubit is observed to be 99.2 %, whereas for a seven-spin E-qubit the measured fidelity is $\approx 99.9$ %. Even at a sub-$K$ temperatures (10 $mK$, 100 $mK$) E-qubit consistently exhibits higher fidelity than a single spin qubit. Moreover, at 6 $K$, the $\pi_x$ gate fidelity for a five-spin ensemble qubit surpasses 99 %, whereas the same operation on a single-spin qubit has a fidelity of only 95.2 %. This indicates that robust single-qubit $\pi_x$ gate operations can still be performed for spin E-qubits in the presence of thermal noise at 6 $K$. In other words, gate operation on E-qubits is more robust than on single-spin qubits in the presence of thermal noise.

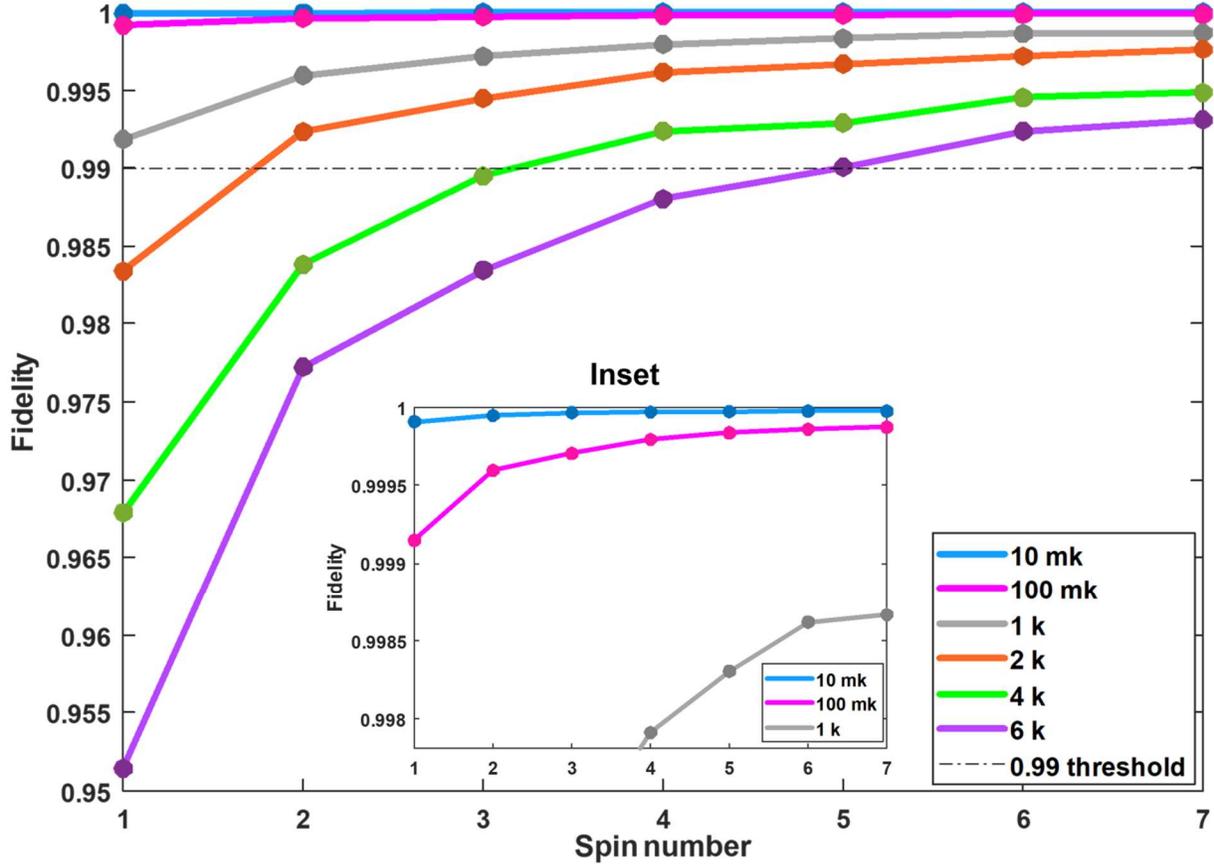

**Figure 5. Relations between average $\pi_x$ gate fidelity and the number of spins in the spin ensemble.** The exchange coupling strength between neighboring spins in E-qubit is considered $7.15 \times 10^{-22}$ $J$ per site. For a single-spin qubit, the fidelity is 95.14 % at 6 $K$ (in violet line). As the number of spins increases, the fidelity rises, with more than 99 % fidelity observed for ensembles consisting of only five -spins at 6 $K$. The inset plot shows the fidelity at 10 $mK$ and 100 $mK$ providing a clear depiction of the improvement in fidelity with increasing spin number in the ensemble.

**Spin echo.** The exchange coupled E-qubit exhibits a longer coherence time compared to a single-spin qubit. The coherence times of E-qubit has a linear relationship with qubit size, as demonstrated in Figure 6(c). To estimate the coherence time $T_2$ for a particular qubit, the expected value of the spin projection, $\rho_z$ is measured at different time delays using a Hahn-echo sequence[25,26]. The Hahn-echo sequence is a combination of $\frac{\pi_x}{2}$ and $\pi_x$ propagators as well as the evolution operator in a superposition state as shown below:

$$U_{Hahn} = e^{\frac{i}{2}(\frac{\pi_x}{2}\Sigma_j S_{xj})} e^{[(\frac{i}{\hbar})(\gamma_e \Sigma_j B_{zj} S_{zj}) + H_{noise}]\tau} e^{[\frac{i}{2}(\pi_x)]} e^{[(\frac{i}{\hbar})(\gamma_e \Sigma_j B_{zj} S_{zj}) + H_{noise}]\tau} e^{\frac{i}{2}(\frac{\pi_x}{2})} \quad (1)$$

To measure the coherence time $T_2$ in a superposition state, the control field pulse amplitude is set to $17.8\ mT$, which is significantly high to achieve a shorter spin modulation time for $\frac{\pi_x}{2}$ $(1\ ns)$ and $\pi_x$ $(2\ ns)$ gates, ensuring minimal perturbation of noise during qubit state transitions. The steps to extract decoherence information in the Hahn-echo sequence are as follows:

i. The qubits are initialized in the ground state $|0\rangle$, followed by the application of a $\frac{\pi_x}{2}$ pulse to transfer the spin state from ground state $|0\rangle$ to the superposition state $|-\rangle_y$.

ii. The system is allowed to evolve for a duration $\tau$ in the presence of noise and field inhomogeneity. During this evolution, dephasing changes the spin's state from pure state to a mixed state.

iii. A $\pi_x$ gate is applied for a short period of time $(2\ ns)$ to invert spin superposition state in the transverse plane.

iv. The system evolves again for the same duration $\tau$ in the presence of noise. During this step, decoherence due to field inhomogeneity is canceled out as spins rephase, producing an echo.

v. A final $\frac{\pi_x}{2}$ pule is applied to rotate spin state from superposition state back to ground state $|0\rangle$.

The degree of decoherence is quantified by the loss in expected value of the spin state, $\rho_z$. In this simulation, a temperature of $2\ K$ with a damping factor $0.01$ is considered. As represented in Figure 6(c), the coherence time is proportional to the number of spins in the ensemble. Coherence-time also has a linear relationship with the damping factor and temperature as described in the methodology part.

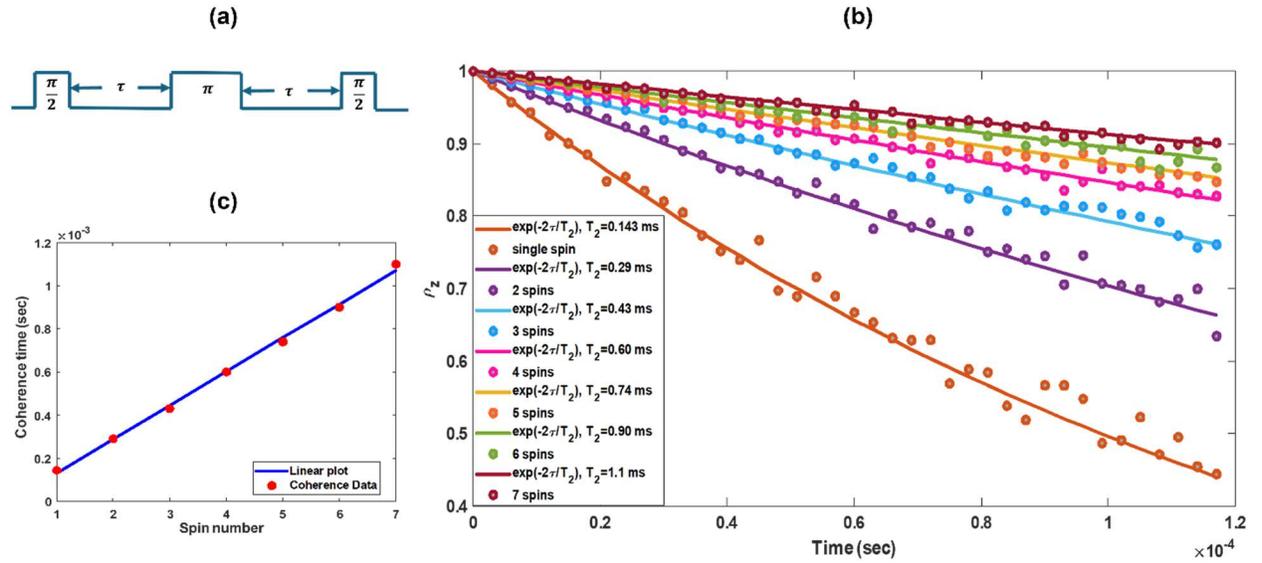

**Figure 6. Comparison of coherence time for different qubit sizes. (a).** The Hahn-echo sequence. Hahn-echo sequence is performed considering temperature $2\ K$ and damping factor $0.01$. **(b** Coherence in terms of expected value $\rho_z$ as function of time delay. **(c).** The linear relation of coherence time with spin number of qubits.

In optically addressable chromium $(IV)$ ion-based molecular qubits embedded in an engineered host matrix with a non-isostructural design, coherence time can be extended to more than $10\ \mu s$ at liquid Helium temperature[27]. In another study, an ensemble of Vanadium (IV)-based molecular spins in a rigid lattice, exhibited a coherence time $T_2$ of $68\ \mu s$ at around $5\ K$ and $1\ \mu s$ at around room temperature[28]. In our study, we demonstrated that a seven spin E-qubit can achieve a coherence time of about $1.1\ ms$ at $2\ K$. Compared to other ensemble spin-based systems, our exchange-coupled E-qubit system offers extended coherence times in the millisecond range, thanks to its intrinsic noise mitigation capability, without requiring carefully shaped pulses or chemical modifications to the material.

In qubit systems, various sources of decoherence exist, including dipolar interaction while we've showed decoherence due to thermal effects causing phonon mediated decoherence from spin-lattice coupling. Although the spin system might experience dephasing due to dipolar interaction, this effect can be mitigated by various suppression schemes. One such approach involves tilting and rotating the geometric plane of the spin ensemble at the magic angle relative to the Zeeman field direction, which can effectively suppress dipolar interaction in the presence of a high Zeeman field[29,30]. Moreover, the use of dynamical decoupling techniques through proper engineering of pulse sequence may help increase the coherence time to some extent[31,32].

There are three $3d$ transitional elements with partially filled d orbital, such as Fe (26), Co (27) and Ni (28) which exhibit ferromagnetic exchange with high curie points ($1043\ K, 1400\ K,$ and $627\ K$ respectively), although the $3d$ electrons remain itinerant at very low temperatures. Some 4f elements with localized electrons that exhibit ferromagnetic behavior such $Gd^{3+}$, $Tb^{3+}$, $Dy^{3+}$, $Ho^{3+}$ etc. have saturation magnetization at 0 k of $7.8\mu_B, 9.7\mu_B, 10.5\mu_B, 10.6\mu_B$ and Curie temperature of $289\ K, 218\ K, 90\ K$ and $20\ K$, respectively[33]. These ions can be deposited using electron vapor deposition or sputtering methods on a MgO substrate to form HCP structure on a $2D$ plane, which can act as ensemble spin.

**Performance of exchange-coupled E-qubit in presence of field inhomogeneity**. In non-ideal case, the magnetic field applied for quantum operations on spin qubit is not perfectly homogenous due to field fluctuations and divergence affecting the coherent rotation of spins. To suppress the decoherence, several techniques used in magnetic resonance for shaping the control pulse have been applied[26,32] such as Knill pulse-a composite pulse generally designed to operate with high fidelity against Zeeman and control field fluctuation[17]. The question arises as to whether the addition of ferromagnetic exchange coupling can also help reduce the effects of field inhomogeneity.

Although it is possible to proximally control an ensemble of spins with a minimal divergence of the global microwave magnetic field, individual spins might interact with varying strengths of microwave and Zeeman fields due to a spatial field gradient, thereby reducing the overall fidelity of the ensemble spin system. We have shown that ferromagnetic exchange interaction acts as an inherent fault mitigator by preventing the dephasing of individual spins.

For this simulation we consider an E-qubit consisting of seven spins in a hexagonal configuration with a central spin, as shown in Figure 2. The bar plot in Figure 7(a) shows the distribution of Zeeman field over the E-qubit, where a 0.5 % field gradient is assumed in $x$ direction. The maximum field of $71.36\ mT$ is applied on peripheral spins while the minimum field of $70.64\ mT$ acts on the central spin. The bar plot in Figure 7(b) shows the distribution of the control field, which has a 10% field gradient in the x-direction, with a maximum amplitude of $1.78\ mT$ on the peripheral spins and a minimum of $1.424\ mT$ on the central spin. The Zeeman field inhomogeneity for each non-interacting spin of the ensemble results in variations in the Larmor precession frequency; in other words, non-interacting spins are off resonance. Under the influence control field non-homogeneity, spins flip at different instants of time as the inhomogeneous control field affects Rabi frequency of individual spins. Figure 7(c) shows fidelity of individual uncoupled spins after a $\pi_x$ operation, expressed as probability of reaching the $|1\rangle$ state. Variation in fidelity, as shown in Figure 8, implies that the inhomogeneity of the control microwave field amplitude changes the overall Rabi oscillation frequency of

E-qubit to $47.281\ MHz$ with a spin-ensemble flip time $21.15\ ns$. Even after tuning the microwave field frequency to the corresponding mean value of Zeeman field and adjusting the pulse width to $21.15\ ns$, as shown in Figure 8, only $85.37\ \%$ fidelity is achieved for the uncoupled ensemble qubit. In contrast exchange coupled E-qubit offers enhanced gate fidelity. As shown in Figure 7(d), the fidelity of the exchange coupled E-qubit reaches $99.99\ \%$ fidelity. Thus, exchange couple E-qubit can act as an intrinsic error mitigator against field inhomogeneity.

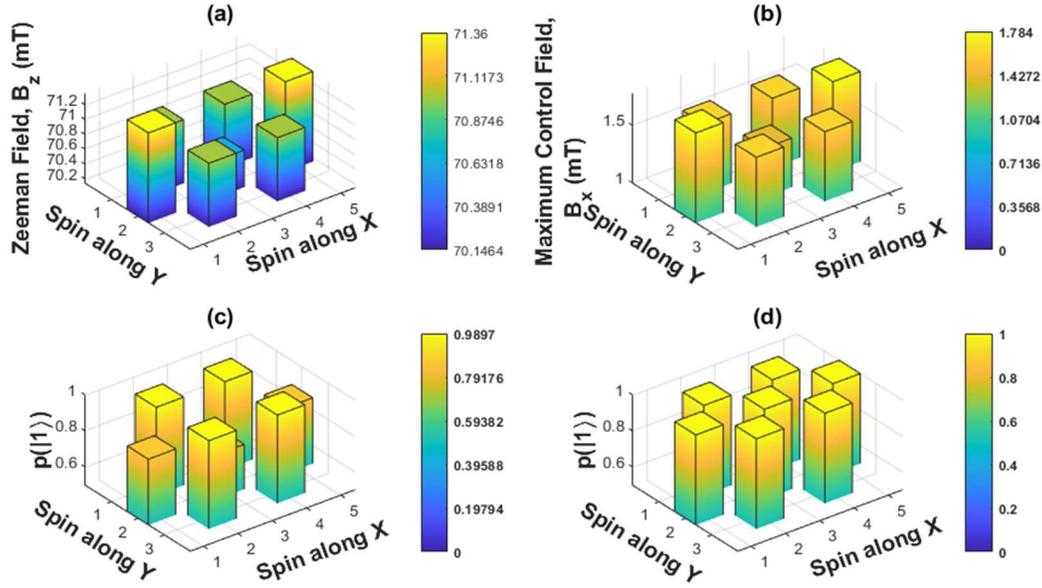

**Figure 7. The effect of field inhomogeneity and exchange interaction on the seven-spin ensemble.** The figure illustrates the robustness of E-qubit consisting of seven ferromagnetically exchange-coupled spins during a quantum $\pi_x$ gate operation under conditions of field inhomogeneity. **(a)** The spatial inhomogeneity of the static Zeeman field, represented by the variation in Larmor frequency. Here, we consider a $0.5\ \%$ gradient in the Zeeman field, corresponding to a frequency variation from $2.00\ GHz$ to $1.98\ GHz$. **(b)** The spatial inhomogeneity of the control field with a $10\ \%$ gradient, ranging from $1.4282\ mT$ to $1.784\ mT$. **(c)** The fidelity of individual non-interacting spins. The fidelities are expressed as the probability $p(|1\rangle)$ of an individual spin reaching the $|1\rangle$ state, with a wide variation observed-from $98.97\ \%$ to $74.87\ \%$. **(d)** The improved fidelity state of the seven spins with the present ferromagnetic Heisenberg exchange. This exchange interactions effectively lock the spin altogether resulting in $99.99\ \%$. fidelity retainment of each spin.

**Ramsey interferometry to investigate $T_2^*$**

We investigate dephasing time due to field inhomogeneity using a Ramsey sequence on both the uncoupled E-qubit and the exchange coupled E-qubit. The propagator of Ramsey sequence[34] is as follows:

$$U_{Ramsey} = e^{\frac{i}{2}(\frac{\pi_x}{2})\sum_j S_{xj}} e^{[(\frac{i}{2})(\gamma_e \sum_j B_{zj} S_{zj}) + H_{noise}]2\tau} e^{\frac{i}{2}(\frac{\pi_x}{2})} \quad (2)$$

The procedure of Ramsey sequences is as follows

i. The E-qubit is initialized in the ground state $|0\rangle$,
ii. A $\frac{\pi_x}{2}$ gate is applied to transform the state from pure state to superposition state,

iii.  The qubit is allowed to undergo Larmor precession for duration $\tau$.
iv.  Another $\frac{\pi_x}{2}$ gate is applied to rotate the spin from transverse direction to again longitudinal direction.

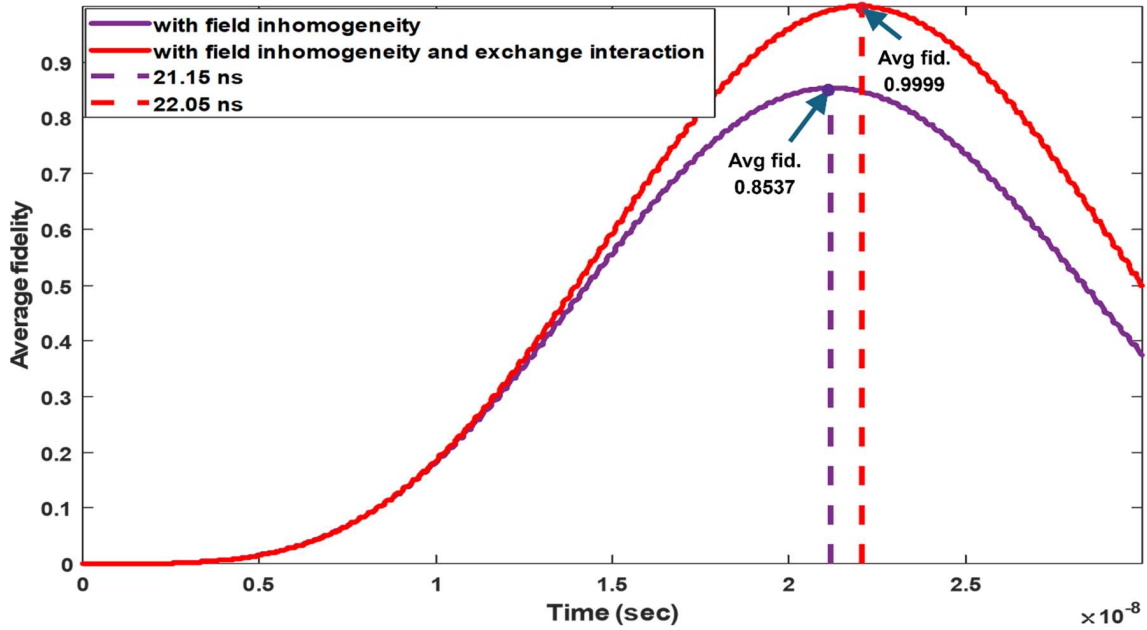

**Figure 8. The plot of the average fidelity versus time for the seven-spin ensemble**. Field inhomogeneity shifts the Rabi oscillation frequency, (violet line), where a maximum fidelity of non-interacting spin ensemble is observed at $21.15\ ns$. The graph also illustrates the decrease of maximum of average spin's fidelity resulting from the field inhomogeneity of the uncoupled spins, while the red line exhibits that fidelity remains above $99.99\ \%$ due to the exchange interaction between neighboring spins.

At the precession period, due to the Zeeman field inhomogeneity, the spins in the ensemble do not precess coherently rather at varying frequencies, resulting in dephasing and a reduction in the Bloch vector magnitude following a Gaussian envelope $\exp(-\tau^2/T_2^*)$. Although this issue can be addressed using dynamical decoupling techniques with composite pulses (e.g., Hahn-echo) with some additional complexities in the quantum control operation, our approach has a simple solution to this problem. We demonstrate that the issue can be resolved by adopting an exchange-coupled E-qubit with inherent dephasing-mitigation capability. In our simulation, the coherence time, $T_2^*$ for E-qubit of uncoupled spins is found to be $34.6\ ns$ as shown in Figure 9, that's inversely proportional to the average detuning in resonant frequency. In this case we consider only $0.5\ \%$ Zeeman field gradient, with a maximum Zeeman field of $71.36\ mT$ and corresponding microwave frequency of $2\ GHz$.

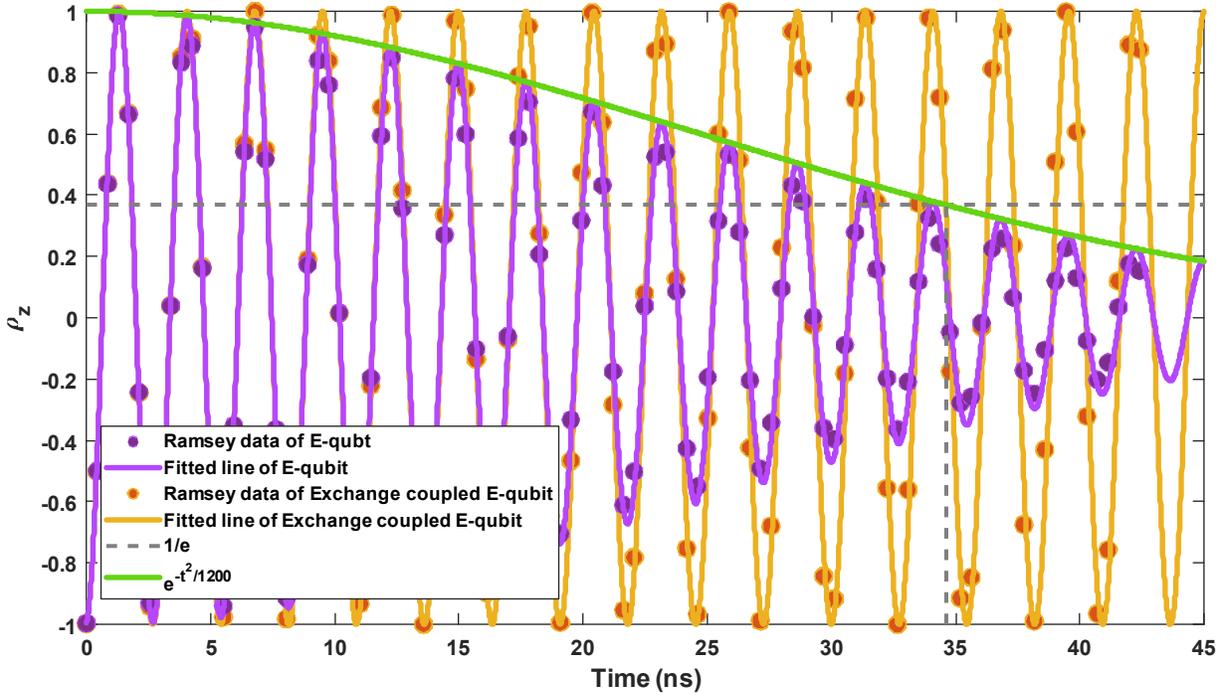

**Figure 9**. Ramsey sequence with and without exchange-coupled E-qubit. Without exchange coupling, the spins exhibit rapid dephasing with rate proportional to the detuning frequency of spins. The calculated $T_2^*$ for non-interacting spin seven-spin E-qubit is $\sqrt{1200} \approx 34.6$ ns. With help of exchange coupling all spin coherently rotate. No decay is visible in case of exchange couple E-qubit.

## Conclusion

In conclusion, this study demonstrates that a ferromagnetically coupled spin ensemble provides a robust qubit framework, exhibiting better performance than single-spin qubit. It enables at least an order of magnitude improvement in fidelity at the same temperature. Alternatively, it allows for operation at temperatures up to ten times higher while maintaining a similar fidelity level. At temperature above $1\,K$, an exchange coupled ensemble qubit consisting of only a few spins can achieve more than $1\,ms$ coherent time which can be further extended linearly by increasing the number of spins. Furthermore, exchange interaction prevents the spins dephasing caused by field inhomogeneity enabling a fault tolerant quantum gate operation in the presence of an inhomogeneous field. With higher fidelity and longer coherence times of exchange coupled ensemble-based qubits enhance computational reliability, improve scalability in quantum circuits and offer a better alternative for advanced sensing applications.

## Methods

The ensemble of spins acting as a qubit can be manipulated using a microwave field as a single $\pi_x$ gate according to the microwave resonance experiment method. Let's consider the field Hamiltonian of the ensemble system for the static magnetic field, $B_0$ along the z-axis and the time-varying field, $B_1$ along the x-axis is given by:

$$H_1 = -\sum_i \gamma_e B_0 \hat{S}_{zi} - \sum_i \gamma_e B_1(t) \hat{S}_{xi} = -\sum_i \omega_0 \hat{S}_{zi} - \sum_i \omega_x(t) \hat{S}_{xi} \qquad (3)$$

where, $\hat{S}_{xi} = \frac{\hbar}{2}\left(\sigma_x \otimes ..I_2^{i^{th}}.. \otimes \sigma_x\right)$ and $\hat{S}_{zi} = \frac{\hbar}{2}\left(\sigma_z \otimes ..I_2^{i^{th}}.. \otimes \sigma_z\right)$

For ease of understanding, the operator can be normalized as follows,

$$\hat{S}_{xi} = \left(\sigma_x \otimes ..I_2^{i^{th}}.. \otimes \sigma_x\right) \qquad (4\ a)$$

$$\text{and } \hat{S}_{zi} = \left(\sigma_z \otimes ..I_2^{i^{th}}.. \otimes \sigma_z\right) \qquad (4\ b)$$

An isotropic exchange can be described as a diagonal tensor

$$H_{exch} = \sum_{i,j}\{S_i.J.S_j\}$$

$$= \sum_{i,j}\left\{[S_{xi}\ S_{yi}\ S_{zi}]\begin{bmatrix}J_x & 0 & 0\\0 & J_y & 0\\0 & 0 & J_z\end{bmatrix}\begin{bmatrix}S_{xj}\\S_{yj}\\S_{zj}\end{bmatrix}\right\} \qquad (5)$$

where $S_i$ is $i^{th}$ spin and $S_j$ is a nearby spin of $i^{th}$ spin $J_x = J_y = J_z$

In case of field inhomogeneity, the applied field Hamiltonian can be written as

$$\begin{aligned}H_1 &= -\frac{\hbar}{2}\sum_i \gamma_e B_{oi}\hat{S}_{zi} - \frac{\hbar}{2}\sum_i \gamma_e B_i(t)\hat{S}_{xn}\\&= -\frac{\hbar}{2}\sum_i \omega_{0i}\hat{S}_{zi} - \frac{\hbar}{2}\sum_i \omega_{xi}(t)\hat{S}_{xi}\end{aligned} \qquad (6)$$

where $\omega_{xi}(t) = B_i(t)\cos(\gamma_e \overline{B_{oi}})t$ , $\overline{B_{oi}} = \frac{1}{N}\sum_i B_{oi}$, $N$ is the number of spins in an ensemble

Pulse width time is directly dependent on the control field $B_i(t)$ amplitude and the frequency can be considered from the mean value of the Zeeman field.

In the case of thermal noise modeling

$$H_{th} = -\sqrt{\frac{2\gamma_e\mu_0 k_B T\alpha}{\mu_e \Delta t}}\sum_i(G_{0,1}^{\ x}(i,t)\hat{S}_{xi} + (G_{0,1}^{\ z}(i,t)\hat{S}_{zi}) \qquad (7)$$

$G_{0,1}^{x,y}$ is Gaussian random distribution with mean 0 and variance is $\sqrt{\frac{2\gamma_e\mu_0 k_B T\alpha}{\mu_e \Delta t}}$ 35 where, $\Delta t$ is time step as small as possible to maintain the good stochasticity, $\alpha$ is damping constant, $\mu_e$ is the magnetic moment of the individual spin, $T$ is temperature, $\mu_0$ magnetic permeability in vacuum.

Combining equations (2), (3), (4), and (5), Total Hamiltonian can be written as follows

$$H_{total} = H_1 + H_{exch} + H_{th} \qquad (8)$$

The ensemble of spins is initialized as $\rho(0) = |00..0\rangle$ state as parallel to the Zeeman field $B_0$

The state term can be written in terms of density matrix as follows

$$\rho(t) = \left\{\frac{1}{2}(I + \sigma_z)\right\}^{\otimes N} \qquad (9)$$

where $\rho(t)$ is the total density matrix of $N$ spins.

Differentiating the density matrix with respect to time, we can find the evolution of the density matrix. From Liouville Von Neumann's equation as follows

$$\frac{\partial \rho(t)^n}{\partial t} = -i[H_{total}(t), \rho(t)] \tag{10}$$

Solution of Eq. (10) can be given in terms of time-ordered exponentials

$$\rho(t) = U\rho(0)U^\dagger \tag{11}$$

where $U_1 = \tau \exp(-i \int_0^t [H_{total}] dt')$

To get spin dynamics of individual spin, the expected value of each spin along $\hat{x}$, $\hat{y}$ and $\hat{z}$ axis can be calculated in each time step as follows

$$\rho_{xi}(t) = Trace(\rho(t).S_{xi}) \tag{12 a}$$

$$\rho_{yi}(t) = Trace(\rho(t).S_{yi}) \tag{12 b}$$

$$\rho_{zi}(t) = Trace(\rho(t).S_{zi}) \tag{12 c}$$

**Fidelity calculation.** For a multilevel system, the average fidelity of a noise-implemented gate can be calculated by comparing the final states generated by the gate's propagator map and the Ideal unitary propagator.

Fidelity between two states $\rho$ and $\eta$ be described as follows[22,23]

$$F(\rho, \eta) = \left( Trace\left( \sqrt{\rho^{\frac{1}{2}} \eta \rho^{\frac{1}{2}}} \right) \right)^2 \tag{13}$$

where $\eta$ can be considered as the qubit's final state comes from its initial arbitrary state by $\pi_x$ operation by the Unitary propagator of the ideal gate. Ideal propagators can be regarded as Unitary $U = \exp\left(\frac{i}{2}\pi\sigma_x\right)$. In this paper, for simplification of calculation, each spin's initial state is considered as $|0\rangle$ instead of an arbitrary state so that after $\pi_x$ operation, the final state should be $|1\rangle$.

In case of noisy implementation of gate operation, the final state of individual spin will not be perfectly in $|1\rangle$ rather spin vector will have azimuthal and polar angle in the Bloch sphere is $|\psi\rangle = \cos\frac{\theta}{2}|0\rangle + e^{i\phi}\sin\frac{\theta}{2}|1\rangle$. The density matrix representation of the final state is

$$\rho = \begin{bmatrix} \cos^2\frac{\theta}{2} & \frac{e^{i\phi}\sin\theta}{2} \\ \frac{e^{-i\phi}\sin\theta}{2} & \sin^2\frac{\theta}{2} \end{bmatrix} \tag{14}$$

In terms of spinor vector representation, it will be, $\rho = \frac{1}{2}(I + \vec{M}.\vec{\sigma})$, where $I$ is the identity matrix and $\vec{M}$ is spinor vector in the Bloch sphere, and $\vec{\sigma}$ is all Pauli matrices.

Let's consider $\rho$ is a positive semidefinite, Hermitian, and Normalized matrix. To calculate the fidelity of a two-level system $\rho^{\frac{1}{2}}$ can be computed using the diagonalization method considering the $\rho$ is vivacious semi-definite matrix that means the eigenvalue of $\rho$ is positive. If $\lambda_1$ and $\lambda_2$ are eigenvalues and $v_1$ and $v_2$ are the eigenvectors (column vectors) of the density matrix $\rho$

$$D = \begin{bmatrix} \sqrt{\lambda_1} & 0 \\ 0 & \sqrt{\lambda_2} \end{bmatrix} \tag{15 a}$$

and $P = [v_1\ v_2],$ (15 b)

thus, $\rho^{\frac{1}{2}} = P D^{\frac{1}{2}} P^{-1}$

$$\rho^{\frac{1}{2}} = \rho \begin{bmatrix} \cos^2\frac{\theta}{2} & \frac{e^{i\phi}\sin\theta}{2} \\ \frac{e^{-i}\sin\theta}{2} & \sin^2\frac{\theta}{2} \end{bmatrix} \tag{16}$$

If the Initial state of each spin is $|0\rangle$ after ideal $\pi_x$ operation final state will be $|1\rangle$

So that, Ideal density state of an individual spin is after $\pi_x$ flip will be $\eta = \begin{bmatrix} 0 & 0 \\ 0 & 1 \end{bmatrix}$

Substituting the term of $\rho^{\frac{1}{2}}$ and $\eta$ in equation (13), the fidelity of an individual can be calculated as follows

$$F(\rho, \eta) = \left( Trace \left( \begin{bmatrix} \cos^2\frac{\theta}{2} & \frac{e^{i\phi}\sin\theta}{2} \\ \frac{e^{-i}\sin\theta}{2} & \sin^2\frac{\theta}{2} \end{bmatrix} \begin{bmatrix} 0 & 0 \\ 0 & 1 \end{bmatrix} \begin{bmatrix} \cos^2\frac{\theta}{2} & \frac{e^{i\phi}\sin\theta}{2} \\ \frac{e^{-i\phi}\sin\theta}{2} & \sin^2\frac{\theta}{2} \end{bmatrix} \right)^{\frac{1}{2}} \right)^2$$

$$F(\rho, \eta) = \sin^2\frac{\theta}{2} \tag{17}$$

In ensemble of spins each spin typically is considered as 2 level system with energy levels split by Zeeman field and the splitting allows the spins in up and down states. The measurement of spin polarization can be conducted by computing the statistical average of all spins state. Although the ensemble exhibits collective entity as single qubit, each individual spin maintains its spin state property as a two-level system. In our computational analysis, obtaining the reduced density matrix for individual spin involves taking partial trace over the rest of the system. Comparing each reduced density state with ideal state provides each individual spin fidelity. In the case of an ensemble with $N$ spins, to get information on the 2-level system of each spin, we take the partial trace of the total density matrices of $2^N$ dimensional Hilbert space to find the reduced density matrix[37] of dimension 2.

$$\rho_i = Trace_{1,2,\ldots i-1, i+1\ldots N}(\rho_{123\ldots N})$$

$$\rho_i = \sum_{s_1 s_2 \ldots s_{i-1} s_{i+1} \ldots s_7} \langle s_1 s_2 \ldots s_{i-1} s_{i+1} \ldots s_7 | \rho_{123\ldots N} | s_1 s_2 \ldots s_{i-1} s_{i+1} \ldots s_7 \rangle \tag{18}$$

Here, $\rho_i$ is individual spin density state and $\rho_{123\ldots N}$ is total density state of N spins in the ensemble.

$\rho_{123\ldots N} = \rho_1 \otimes \rho_2 \otimes \ldots \otimes \rho_N$ in $2^N$ dimensional Hilbert space.

As E-qubit consisting of large ensemble spins requires less iterative quantum measurement as expected value of spin ½ qubits state from the ensemble can be found by a single shot 2 levels measurement after quantum operation, whereas single spin requires repetitive operation/measurement to get the same expected result[11]. Let's assume $P$ is the number measurement required for single-spin qubit, if $N$ is number of spins in an ensemble, measurement required to get similar statistical information will be $\frac{P}{N}$. In the case of positive operator valued measurement, with $\frac{P}{N}$ number of iterations, fidelity can be calculated by comparing individual spin's final state in $N$-spin ensemble with ideal state $|1\rangle$.

$$F(\rho,\eta) = \left(\frac{1}{P}\right)\sum_1^P \sum_{i=1}^N p_i \sin^2\frac{\theta_i}{2}|1\rangle\langle 1|) \tag{19}$$

**Interpretation for longer $T_2$ in case of exchange coupled E-qubit**

In Bloch sphere, spin state is regarded as a Bloch vector, the density matrix of single-spin state can be written as

$$\rho = \frac{1+M.\sigma}{2} \tag{20}$$

as $M$ is a Bloch vector and $\sigma$ is Pauli operator.

In case of magnetic resonance, Liouville Von Neumann equation can be written as Bloch vector precession equation:

$$\dot{\vec{M}} = \vec{M} \times \vec{B} \tag{21}$$

Magnetic field term is $\vec{B(t)} = \vec{B_0} + \vec{b_{noise}}$ with corresponding frequency terms $\omega(t) = \omega_L + \delta\omega(t)$. $\delta\omega(t)$ is related to Gaussian random noise. with mean 0 and standard deviation[24], $\sigma = \sqrt{\frac{2\gamma_e \mu_0 k_B T \alpha}{\mu_e \Delta t}}$

$m_+(t) = e^{i\phi(t)} m_+(0)$, where accumulated phase, $\phi(t) = \int_0^t [\omega_L + \delta\omega(t')] dt'$ and $m_+ = (M_x + iM_y)/M$

In Hahn-echo sequence, before applying the $\pi$ gate, the spin is allowed to evolve in the transverse plane for a duration $\tau$ with Larmor precession. During the period, dephasing may occur due to Zeeman field inhomogeneity or frequency detuning. After applying the $\pi_x$ gate within a short period, the spin is allowed to evolve for a duration $\tau$. At this stage, the spins rephase, effectively reversing the time evolution.

So that, Stochastic part of phase change for a duration $2\tau$ should be,

$$\phi_{Hahn} = \phi_1 - \phi_2 = \int_0^\tau \delta\omega(t')dt' - \int_\tau^{2\tau} \delta\omega(t')dt' \tag{22}$$

Although dephasing due to field inhomogeneity can be canceled out, stochastic noise still leads to decoherence. Since random phase is a sum of uncorrelated correlated noise, the central limit theorem suggests that phase fluctuations should follow a Gaussian distribution.

$$p(\varphi) = \frac{1}{\sqrt{(2\pi\langle\varphi^2\rangle)}} \exp\left(-\frac{\varphi^2}{2\langle\varphi^2\rangle}\right) \tag{23}$$

Expected value of transverse Bloch vector will be

$$\langle e^{i\phi(t)}\rangle = \int p(\varphi) e^{i\varphi} = e^{-\frac{\langle\varphi^2\rangle}{2}} \tag{24}$$

Variance of total phase accumulation is

$$\langle\Delta\phi_{Hann}(t)^2\rangle = \langle(\phi_1 - \phi_2)^2\rangle = \langle\phi_1^2\rangle + \langle\phi_2^2\rangle - 2\langle\phi_1\rangle\langle\phi_2\rangle$$

$$\langle\phi_1^2\rangle = \langle\phi_2^2\rangle = \int_0^\tau dt_1 \int_0^\tau \langle\delta\omega(t_1)\delta\omega(t_2)\rangle dt_2 = \sigma^2 \int_0^\tau dt_1 \int_0^\tau dt_2\, \delta(t_1 - t_2) dt_2 = \sigma^2\tau$$

$$\langle\phi_1\phi_2\rangle = \int_0^\tau dt_1 \int_\tau^{2\tau} \langle\delta\omega(t_1)\delta\omega(t_2)\rangle dt_2 = \frac{\sigma^2}{2} \int_0^\tau dt_1 \int_0^\tau dt_2\, \delta(t_1 - t_2) dt_2,$$

As $\delta(t_1 - t_2)$ ensures no overlap between the two intervals, $t_1 \in [0,\tau]$ and $t_2 \in [\tau, 2\tau]$

$$\langle\phi_1\phi_2\rangle = 0$$

Total phase variance is

$$\langle \Delta\phi_{Hahn}(t)^2 \rangle = \sigma^2\tau + \sigma^2\tau - 0 = 2\sigma^2\tau \tag{25}$$

From substituting the random phase term in equation (24) we get,

So that, $\quad \langle e^{i\phi(2\tau)} \rangle = e^{-\sigma^2\tau} = e^{-\frac{t}{T_2}},$

Thus, $\quad T_2 = \frac{1}{\sigma^2} = \frac{\mu_e \Delta t}{2\pi\gamma_e\mu_0 k_B T \alpha} \tag{26}$

That means coherence Time $T_2$ is inversely proportional to Temperature and Damping Factor.

For multiple spins system, let's consider two spins coupled with ferromagnetic exchange. The initial state of the spin system is triplet state $T_-$ i.e $|00\rangle$

According to the Hahn-echo sequence after applying the mw pulse to implement $\frac{\pi}{2}$ gate in resonant condition, the spins will rotate from $T_-$ state to superposition state. Due to strong ferromagnetic exchange coupling spins will remain in triplet state. The final state in the transverse plane will be

$$|\psi\rangle = \frac{1}{\sqrt{2}}(|0\rangle - i|1\rangle) \otimes \frac{1}{\sqrt{2}}(|0\rangle - i|1\rangle)$$

$$= \frac{1}{2}(|00\rangle - i(|01\rangle + |10\rangle) - |11\rangle)$$

$$= \frac{1}{2}(T_- - i\sqrt{2}T_0 - T_+) \tag{27}$$

$|\psi\rangle$ is a degenerate triplet state with no singlet terms

In the symmetric triplet states the two spins act as one collective spin. During the evolution of dephasing and rephasing before and after $\pi$ gate respectively, spins suffer decoherence due to noise. The effective noise is the average of two individual spins.

For spin precession duration t, Larmor frequency, $\omega_L = \gamma_e B_0$, and frequency fluctuation, $\delta\omega_i(t') = \gamma_e b_{i,noise}$ phase accumulated is determined by

$$\varphi = \int_0^t \left[\omega_L + \frac{\delta\omega_1(t') + \delta\omega_1(t')}{2}\right] dt' \tag{28}$$

So that the effective noise will be, $\delta\omega_{eff}(t) = \frac{\delta\omega_1(t') + \delta\omega_1(t')}{2}$

with variance, $\langle \delta\omega_{eff}(t)^2 \rangle = \frac{\sigma^2}{2}$; half of variance compared to single-spin qubit

Phase accumulation after the Hahn-echo is

$$\phi_{Hahn} = \phi_1 - \phi_2$$

$$= \int_0^\tau \delta\omega_{eff}(t')dt' - \int_\tau^{2\tau} \delta\omega_{eff}(t')dt' \tag{29}$$

Total phase variance is $\langle \Delta\phi_{Hahn}(t)^2 \rangle = \frac{\sigma^2\tau}{2} + \frac{\sigma^2\tau}{2} - 0 = \sigma^2\tau \tag{30}$

Variance of the noise in case of thermal field can be written as $\sigma^2 = \frac{2\gamma_e\mu_0 k_B T \alpha}{\mu_e \Delta t}$

So $\langle e^{i\phi(2\tau)}\rangle = \int p(\varphi)e^{i\varphi} = e^{-\frac{\langle\varphi^2\rangle}{2}} = e^{-\frac{\sigma^2\tau}{2}} = e^{-\frac{\tau}{T_2}}, T_2 = \frac{2}{\sigma^2}$

Similarly in case of $N$ number of spins coherence time, $T_2 = \frac{N}{\sigma^2}$ (31)

Substituting the $\sigma^2$ with thermal field fluctuation terms,

$$T_2 = \frac{N\mu_e \Delta t}{\gamma_e \mu_0 k_B T \alpha}$$ (32)

Coherence time $T_2$ is proportional to spin number and inversely proportional to temperature.

## References

1. Monroe, C. *et al.* Programmable quantum simulations of spin systems with trapped ions. *Rev Mod Phys* **93**, 25001 (2021).

2. Pla, J. J. *et al.* High-fidelity readout and control of a nuclear spin qubit in silicon. *Nature* **496**, 334–338 (2013).

3. Cory, D. G., Fahmy, A. F. & Havel, T. F. *Ensemble Quantum Computing by NMR Spectroscopy (NMRquantum ComputingDNA Computingnondeterministic Polynomial-Time Complete)*. vol. 94 www.pnas.org. (1997).

4. Gershenfeld, N. A. & Chuang, I. L. *Bulk Spin-Resonance Quantum Computation. SCIENCE* vol. 275 https://www.science.org (1997).

5. Arute, F. *et al.* Quantum supremacy using a programmable superconducting processor. *Nature* **574**, 505–510 (2019).

6. Chatterjee, A. *et al.* Semiconductor qubits in practice. *Nature Reviews Physics* **3**, 157–177 (2021).

7. Asaad, S. *et al.* Coherent electrical control of a single high-spin nucleus in silicon. *Nature* **579**, 205–209 (2020).

8. Mądzik, M. T. *et al.* Precision tomography of a three-qubit donor quantum processor in silicon. *Nature* **601**, 348–353 (2022).

9. Pezzagna, S. & Meijer, J. Quantum computer based on color centers in diamond. *Applied Physics Reviews* vol. 8 Preprint at https://doi.org/10.1063/5.0007444 (2021).

10. Rugar, A. E. *et al.* Quantum Photonic Interface for Tin-Vacancy Centers in Diamond. *Phys Rev X* **11**, (2021).

11. Niknam, M., Schwartz, R. N. & Bouchard, L.-S. Quantum gates between mesoscopic spin ensembles. *Phys Rev A (Coll Park)* **107**, 32601 (2023).

12. Burkard, G., Ladd, T. D., Pan, A., Nichol, J. M. & Petta, J. R. Semiconductor spin qubits. *Rev Mod Phys* **95**, 25003 (2023).

13. Petit, L. *et al.* Spin Lifetime and Charge Noise in Hot Silicon Quantum Dot Qubits. *Phys Rev Lett* **121**, 76801 (2018).

14. Koppens, F. H. L. *et al.* Driven coherent oscillations of a single electron spin in a quantum dot. *Nature* **442**, 766–771 (2006).

15. Jing, B. & Bao, X.-H. Ensemble-Based Quantum Memory: Principle, Advance, and Application. in *Photonic Quantum Technologies* 433–462 (2023). doi:https://doi.org/10.1002/9783527837427.ch17.

16. Scheuer, J. *et al.* Robust techniques for polarization and detection of nuclear spin ensembles. *Phys Rev B* **96**, 174436 (2017).

17. Niknam, M. *et al.* Quantum control of spin qubits using nanomagnets. *Commun Phys* **5**, (2022).

18. Fahim, M. *et al. Proximal Quantum Control of Spin and Spin Ensemble with Highly Localized Control Field from Skyrmions*.

19. Zu, H., Dai, W. & de Waele, A. T. A. M. Development of dilution refrigerators—A review. *Cryogenics (Guildf)* **121**, 103390 (2022).

20. Veldhorst, M. *et al.* An addressable quantum dot qubit with fault-tolerant control-fidelity. *Nat Nanotechnol* **9**, 981–985 (2014).

21. Peng, S. *et al.* Origin of interfacial perpendicular magnetic anisotropy in MgO/CoFe/metallic capping layer structures. *Sci Rep* **5**, 18173 (2015).

22. Jozsa, R. Fidelity for Mixed Quantum States. *J Mod Opt* **41**, 2315–2323 (1994).

23. Uhlmann, A. The "transition probability" in the state space of a ∗-algebra. *Reports on Mathematical Physics* **9**, 273–279 (1976).

24. Lee, K. J., Park, N. Y. & Lee, T. D. Numerical study of spin relaxation by thermal fluctuation: Effect of shape anisotropy. *J Appl Phys* **89**, (2001).

25. Hahn, E. L. Spin Echoes. *Physical Review* **80**, 580–594 (1950).

26. Levitt, M. H. Composite pulses. *Prog Nucl Magn Reson Spectrosc* **18**, 61–122 (1986).

27. Bayliss, S. L. *et al.* Enhancing Spin Coherence in Optically Addressable Molecular Qubits through Host-Matrix Control. *Phys Rev X* **12**, 31028 (2022).

28. Bader, K. *et al.* Room temperature quantum coherence in a potential molecular qubit. *Nat Commun* **5**, 5304 (2014).

29. Tang, Y., Kao, W., Li, K.-Y. & Lev, B. L. Tuning the Dipole-Dipole Interaction in a Quantum Gas with a Rotating Magnetic Field. *Phys Rev Lett* **120**, 230401 (2018).

30. Ding, S. *et al.* Quantum computation based on magic-angle-spinning solid state nuclear magnetic resonance spectroscopy. *The European Physical Journal B - Condensed Matter and Complex Systems* **24**, 23–35 (2001).

31. Naydenov, B. *et al.* Dynamical decoupling of a single-electron spin at room temperature. *Phys Rev B* **83**, 81201 (2011).

32. Ryan, C. A., Hodges, J. S. & Cory, D. G. Robust Decoupling Techniques to Extend Quantum Coherence in Diamond. *Phys Rev Lett* **105**, 200402 (2010).

33. CHIKAZUMI, S. Physics of Ferromagnetism.

34. Ramsey, N. F. A Molecular Beam Resonance Method with Separated Oscillating Fields. *Physical Review* **78**, 695–699 (1950).

35. Lee, K. J., Park, N. Y. & Lee, T. D. Numerical study of spin relaxation by thermal fluctuation: Effect of shape anisotropy. *J Appl Phys* **89**, (2001).